\documentclass[runningheads,a4paper]{llncs}

\usepackage{amssymb}
\usepackage{graphicx}
\usepackage[caption=false]{subfig}
\usepackage{booktabs}
\usepackage{multirow}

\usepackage{url}
\urldef{\mailsa}\path|{alfred.hofmann, ursula.barth, ingrid.haas, frank.holzwarth,|
\urldef{\mailsb}\path|anna.kramer, leonie.kunz, christine.reiss, nicole.sator,|
\urldef{\mailsc}\path|erika.siebert-cole, peter.strasser, lncs}@springer.com|    

\usepackage{xcolor}

\begin{document}

\mainmatter  % start of an individual contribution

%\title{Lecture Notes in Computer Science:\\Authors' Instructions
%for the Preparation\\of Camera-Ready
%Contributions\\to LNCS/LNAI/LNBI Proceedings}
\title{
Flexible Conditional Image Generation of Missing Data with Learned Mental Maps
}
\titlerunning{Conditional Image Generation of Missing Data with Learned Mental Maps}

\author{Benjamin Hou, Athanasios Vlontzos, Amir Alansary, \\ Daniel Rueckert and Bernhard Kainz}
%index{Hou, Benjamin}
%index{Vlontzos, Athanasios}
%index{Alansary, Amir}
%index{Rueckert, Daniel}
%index{Kainz, Bernhard}
%
\authorrunning{B. Hou et al.}

\institute{
Biomedical Image Analysis Group, Imperial College London, UK
%\url{https://biomedia.doc.ic.ac.uk}
}


\toctitle{Lecture Notes in Computer Science}
\tocauthor{Authors' Instructions}
\maketitle

\begin{abstract}

Real-world settings often do not allow acquisition of high-resolution volumetric images for accurate morphological assessment and diagnostic. In clinical practice it is frequently common to acquire only sparse data (e.g. individual slices) for initial diagnostic decision making. Thereby, physicians rely on their prior knowledge (or mental maps) of the human anatomy to extrapolate the underlying 3D information. Accurate mental maps require years of anatomy training, which in the first instance relies on normative learning, i.e. excluding pathology.  
In this paper, we leverage Bayesian Deep Learning and environment mapping to generate full volumetric anatomy representations from none to a small, sparse set of slices. We evaluate proof of concept implementations based on Generative Query Networks (GQN) and Conditional BRUNO using abdominal CT and brain MRI as well as in a clinical application involving sparse, motion-corrupted MR acquisition for fetal imaging. Our approach allows to reconstruct 3D volumes from 1 to 4 tomographic slices, with a SSIM of 0.7+ and cross-correlation of 0.8+ compared to the 3D ground truth. 

%\keywords{We would like to encourage you to list your keywords within the abstract section}
\end{abstract}

\section{Introduction}
Physical as well as physiological constraints on tomographic image acquisition (e.g. motion) often prohibit the acquisition of high resolution volumetric images that are commonly used for morphological examinations and diagnosis. Acquisition of high resolution images requires a fixed period of time where the patient is asked to remain still, this is often not possible in cases such as fetal imaging. Motion during this period causes scanned slices to become incoherent and corrupt. Long periods of CT scans also impose high levels of exposure to ionising radiation. Single slice or sparse acquisition can often be mentally extrapolated to a 3D mental map by experienced physicians. However, it relies on years of experience and training, thus the need to perform sparse reconstruction arises. In this paper, we address both the need to perform sparse reconstruction, as well as creating mental maps of anatomies. 

Extrapolation of 3D volumes have advantages for tracking and interventional applications. Tracking, e.g. methods such as freehand ultrasound, can benefit the sonographer greatly by providing an extrapolated 3D volume for better spatial reference. Furthermore, iterative image-based motion compensation methods needs a good initial target, which is often not possible to obtain if the subject is awake and constantly in motion during image acquisition. Thus, the need to extrapolate a full 3D volume from very sparse amount of slices is highly desirable.

We leverage Bayesian Deep Learning (BDL) and environment mapping to generate full volumetric anatomy representations derived from none to a few conditioning slices. In contrast to commonly used Conditional Variational AutoEncoders (C-VAE), our model leverages traditional statistical methods where the conditioning variable is not fixed or restricted. This therefore enables us to perform reconstruction of normative structures, extrapolate sparse image acquisition and create mental maps of anatomies. Contrary to previous approaches of sparse reconstruction, as detailed in the related work section below, our method can also produce probabilistic mental representations of the anatomy and anatomical context in question to aid diagnosis and therapy.

\textbf{Related Work:}
Sparse reconstruction of anatomical structures has been the topic of extensive work as a method to reduce cost and, \emph{e.g.}, exposure to ionising radiation for patients and doctors alike. Early approaches included deformable statistical models~\cite{ehlke2013fast} to set a prior to the reconstruction process. More recent approaches have been adopting neural networks and deep learning to perform sparse reconstruction. Cerrolaza et al.~\cite{cerrolaza20183d} uses a hierarchical C-VAE, where given three standard plane views from a 2D ultrasound scan of a fetal brain, to reconstruct the 3D segmentation mask of the fetal skull. Similarly~\cite{wang2017unsupervised} use a Convolutional Auto-Encoder to construct a shared latent space between 2D and 3D images to aid the reconstruction of a 3D image. In addition~\cite{10.1007/11569541_46} perform an inter-domain sparse reconstruction as they perform segmentation of 3D volumes based on 2D sparse data inputs. In the field of natural images~\cite{choy20163d} suggested an iterative technique of refining the 3D reconstruction as the model is given more views. Finally in~\cite{Kunter2009} used stereoscopic reconstruction to achieve 2D to 3D segmentation reconstruction.

\textbf{Contribution:} 
We introduce a method to generate missing slices, via BDL, by sampling from a distribution on the image manifold, which is conditioned on sparse scanned slices as context. We restructure the Conditional BRUNO~\cite{korshunovaconditional} architecture, and train the model to learn a mental map of a specific region of interest or anatomy. The novel aspect of our work is that conditional image generation is not achieved by commonly used C-VAE architectures, but instead through Normalising Flows~\cite{rezende2015variational} and statistical modelling such as Student's t-process. This is applied to generate patient specific dense medical volumes, and evaluated on three different data-sets. To generate patient specific dense medical volumes, we query the model by performing a dense sweep of all possible pose positions, while conditioning the model on sparsely sampled context slices from the patient. The method is evaluated on three data-sets and we demonstrate its application for motion correction in fetal brain MRI.

\textbf{Background:} 
Generative Models are used to model probabilistic distributions, $p(x)$, of a data-set, $X$, such that $x \in X$ in some high-dimensional space $\mathcal{X}$. The model can then be used to generate new samples, such as images, that follows the same probabilistic distribution. New samples are seeded by a latent variable, a vector often denoted $z$ in some high-dimensional space $\mathcal{Z}$, and are sampled according to some Probability Density Function (PDF) $p(z)$. Given a fixed deterministic function, $f(z;\theta)$, parameterised by $\theta$ in some high-dimensional space $\Theta$ and $f:\mathcal{Z} \times \Theta \rightarrow \mathcal{X}$, the aim is to optimise $\theta$ such that samples of $z$ from $p(z)$, and subsequently $f(z;\theta)$, will be similar to $x$  with high probability. Formally, this can then be written as: $p(x) = \int p_{\theta}(x|z)p_{\theta}(z)dz$. 

The most commonly used distribution, also known as the prior, for the latent space is a Gaussian, with a mean of zero and unit variance ($\mathcal{N}(0,1)$). An important component of the modelling process is defining a bijector, which is an invertible fixed transformation function that maps one data space to another. $f$, defined above, can be used to map the complex distribution of the input data space to the $z$ latent space. The distribution of modelled image space can then be written as: $X \sim \mathcal{N}(f(z;\theta),\,\sigma^{2}I)$. As the probability integral is high-dimensional and complex, a neural network can be used to learn $f$. Therefore it is possible to use gradient descent (or any other optimisation technique) to perform $\max_{\theta} \sum_{i} \log(p_{\theta}(x_i))$, which aims to find an optimal set of parameters $\theta$, for the fixed deterministic function $f(z;\theta)$.

Such mappings can be achieved through methods such as Variational AutoEncoders (VAEs)~\cite{DBLP:journals/corr/KingmaW13} or Generative Adversarial Networks (GANs)~\cite{goodfellow2014generative}. The encoder and decoder component of the VAE models the forward and inverse of the bijector function, but are learned separately. In GANs, only the fixed function from latent $z$ space to data space is learned. RealNVP~\cite{DBLP:journals/corr/DinhSB16} and Masked Autoregressive Flow (MAF)~\cite{papamakarios2017masked}, however, are bijectors that are fully invertible. The same weights are used for both forward and inverse transformations.

Conditional Generative Models, such as C-VAEs~\cite{DBLP:conf/nips/SohnLY15}, Generative Query Network (GQN)~\cite{Eslami1204}, or BRUNO~\cite{korshunovaconditional}, generate new samples based on a predefined condition such that for each possible value of $c$ there exists a $p(z)$; $p(x|c) = \int p_{\theta}(x|z)p_{\theta}(z,c)dz$. For this particular task, $c$ is a set of images that are sparsely acquired, and is not bound by quantity. This requires the set of images in the condition being exchangeable, i.e. the joint probability is invariant to permutation of the images. For any permutation, $\pi$; $p(x_{1}, x_{2}, ..., x_{n}) = p(x_{\pi_1}, x_{\pi_2}, ..., x_{\pi_n})$. Random variables are often  independent and identically distributed (iid), and iid random variables are always infinitely exchangeable. However, the converse is not always true, an infinitely exchangeable sequence is not necessarily iid. Bruno de Finetti's theorem therefore states `a sequence of random variables $(x_{1}, x_{2}, ..., x_{n})$ is infinitely exchangeable \emph{iff} for all $n$; $p(x_{1}, x_{2}, ..., x_{n}) = \int \prod_{i=1}^{n} p(x_{i} | \theta) p(\theta) d\theta$. The stochastic process is then defined; $p(x_n | x_{1:n-1}) = \int p_{\theta}(x_n|z)p_{\theta}(z|x_{1:n-1})dz$.

\section{Method}

In the proposed framework, the Conditional BRUNO architecture is restructured and trained to build mental maps of medical volumes using 2D slices, $x$, with their corresponding pose parameter, $v$, that represents the slice's location in 3D space. Similar to a GQN, the data-set is of the form $D = \{(x_n^k,v_n^k)\}$, where $n \in \{1,2,...,N\}$ and $k \in \{1,2,...,K\}$. $N$ is the number of high resolution 3D volumes and $K$ is the number of 2D axial slices of the volumes. During training, $M$ random image-pose pairs, are sampled from a particular volume. Each $m \in M$ is a particular observation, with the collective being denoted as a \textbf{sequence}. $M-1$ observations from the sequence are used as \textbf{contexts}, with the remaining image-pose pair being used as the \textbf{query pose} and \textbf{target image}. 

Each individual context (i.e. image-pose pair) in a sequence is passed through a Conditional RealNVP~\cite{DBLP:journals/corr/DinhSB16}, which is the bijector of the model. The affine coupling layer uses Convolutional Neural Networks (CNN), and learns the mapping of the input image distribution to a Gaussian prior. The mapping is conditioned on the pose, and is made possible by augmenting the input image with the pose vector as an additional input variable. This CNN can be of any structure, for simplicity, a simplified ResNet was used. As the RealNVP component is trained on a Gaussian prior, the output variables should therefore be Gaussian distributed. This can then be modelled by classical statistical methods such as Student's t-distribution, $\mathcal{TP}$. To achieve exchangability, i.e. the conditioning context set is invariant to the number and order of the contexts, a na\"{i}ve approach would be to simply perform sum/average operation similar to~\cite{DBLP:conf/nips/ZaheerKRPSS17}. Alternatively, a Recurrent Neural Network (RNN) update scheme~\cite{korshunovaconditional}, can also be used, where the covariance matrix of the conditional image set is made to be simple (i.e. the diagonal is parameterised by $\mu$, with upper and lower triangle parameterised by $\sigma$), each image therefore has an identical mean and variance to one other. 

\begin{figure}[t]
\centering
\includegraphics[width=\textwidth]{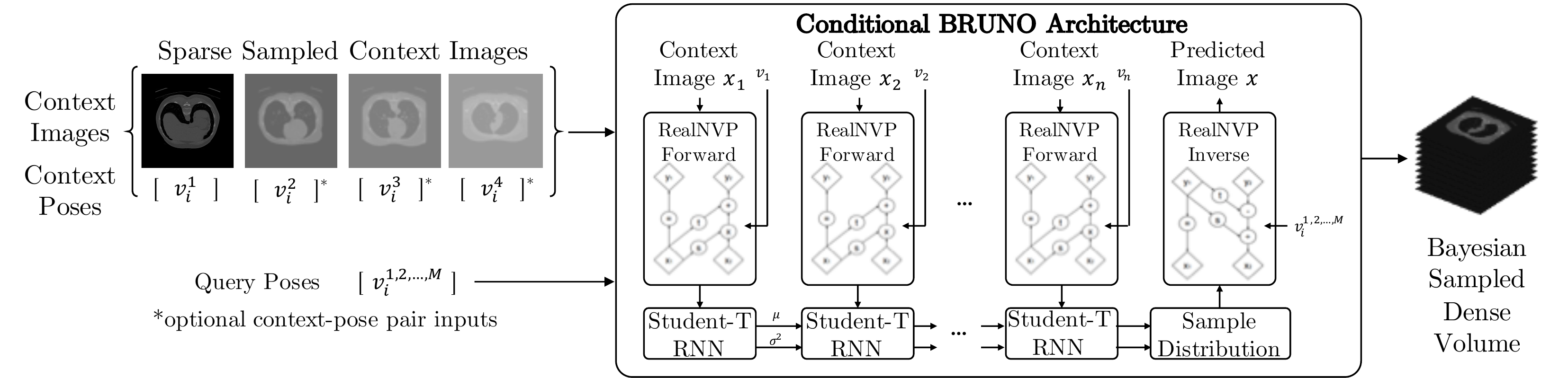}
\caption{BRUNO Architecture for the generation of volumetric anatomical mental maps from very sparse data. Conditioning context images $v_i^{1,2,...,M}$ can be any sample from the observed anatomy in contrast to defined samples in, e.g., C-VAEs.}
\label{fig:bruno-architecture}
\end{figure}

During testing, the number of context image-pose pairs is not required to be the same as the number as used during training, due to the property of exchangeability. Intuitively speaking, as more context images are supplied, the predicted image should look more similar to the target image. Contexts are first passed through the RNN to set up the distribution at the condition by updating the mean and variance, from which the samples can then be drawn. Each sample, drawn from this distribution, is then passed though the inverse Conditional RealNVP, whilst being conditioned on the query pose. 

To create a dense 3D volume from very sparsely sampled 2D slices, a trained BRUNO model is queried with a dense sweep of all possible pose positions within the same Field-of-View (FoV) as training. The sparse sampled 2D slices, along with the corresponding pose, are therefore supplied as contexts for the model. Multiple samples can then be drawn from the trained distribution as possible hallucinations of the missing slice. Alternatively, it is also possible to take the mean image (i.e. average of infinite samples). With no contexts supplied, samples are drawn from the prior distribution. This can be used to create organ and/or volume atlases and for manual model validation, as the trained distribution is an average of all training volumes. Patient specific missing slices and extrapolated anatomy can be generated if context images are supplied. Samples are then drawn from the posterior as the distribution is conditioned on the contexts. 

\section{Experiments and Results}
To validate the trained model, high resolution 3D volumes from the test set are decomposed into individual 2D slices with their corresponding one-hot pose vector. A sparse set of slices (between 1 to 10) are used as contexts. The model is then queried with the pose vectors, the generated slices are then compared to the target image using Cross-Correlation (CC) and Structural Similarity Index (SSIM). CC measures the similarity of pixel-wise intensities, whereas SSIM uses a combination luminance, contrast and structure to assess the image quality. 

The first set of experiments used brain MRI and thorax CT images. 85 healthy brains were selected from the Alzheimer's Disease Neuroimaging Initiative (ADNI) database. These were split 70 for training and 15 for validation, with $K=120$ axial slices extracted from the middle 75\% portion of the brain. The CT images were split 50 for training and 8 for validation, with $K=100$ axial slices extracted from the middle 50\% portion of the scan volume. Both data-sets are isotropic with spacing of $1mm\times1mm\times1mm$. Each 2D slice, $x$, are of size $218\times218$ for MRI brain and $200\times200$ for CT thorax, and are further down-sampled to $64\times64$. An additional isotropic fetal brain MRI data-set, with spacing of $1mm\times1mm\times1mm$, was used for Initial Experiments and Exp2. 270 brains ranging from 40 to 43 Gestational Age (GA) were selected; split 250 for training and 20 for testing. $K=80$ axial slices were then extracted from the middle 65\% of the brain, each 2D slice of size $160\times160$ is further down-sampled to $64\times64$ for training and inference. 

The pose is formulated as one-hot vector of length $K$, to represent the slice number of the scan stack. For all experiments, $M=9$; a sequence of eight image-pose pairs are used as context, with the 1 remaining as query pose and target image. Contexts are randomly sampled from the entire stack during training phase. However, during testing, the contexts are strategically selected so that they cover an approximate even distribution across the scan stack. For each query pose, 100 samples were drawn from the posterior distribution and compared to the target image using SSIM and Cross-Correlation. An average is then taken across all slices for a volume average, and further averaged across all test subjects.

\subsection{Initial Experiments}
To compare with existing baseline models, GQN and C-VAE architectures were used to build mental maps using the Fetal brain data-set. Both GQN and C-VAE models have been trained with 4 context image-pose pairs. 

As official code for the Generative Query Network have not been published, a reputable public reimplementation~\cite{ogroth_2019} was used instead. The implementation has been validated to be correctly functioning, as it has been successfully tested on several, but simple, official GQN data-sets. Performance however was not able to match the results published by DeepMind, as the architecture for the DRAW-LSTM was not disclosed. Only the baseline architecture was used. 

A na\"{i}ve Conditional Variational AutoEncoder was also implemented as a baseline. The architecture follows the structure of a standard U-Net~\cite{ronneberger2015u}, but without skip connections, and with 4 scaled resolutions and 2 convolutional layers at each resolution. It has also been formulated such that the input is a pose vector, the condition is a set of encoded context images, and the output is the generated image. Contexts are first passed through a tower encoder network, same as the GQN, to encode each image-pose pair in latent representation. All Contexts are then averaged together in latent space to maintain order invariance. During inference, the latent vector $z$ is sampled from a unit Gaussian distribution to feed the generator whilst being conditioned on contexts as well as queried pose. 

\begin{figure}[t]

  \centering

  \subfloat[$v$: 10]{\includegraphics[width=1.4cm]{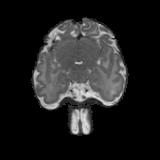}} \hspace{2mm}
  \subfloat[$v$: 30]{\includegraphics[width=1.4cm]{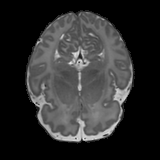}} \hspace{2mm}
  \subfloat[$v$: 50]{\includegraphics[width=1.4cm]{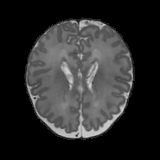}} \hspace{2mm}
  \subfloat[$v$: 70]{\includegraphics[width=1.4cm]{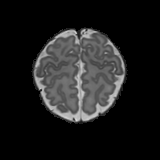}} \hspace{2mm} \\
  
  \subfloat{\includegraphics[width=1.4cm]{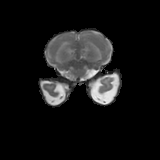}}  \hfill
  \subfloat{\includegraphics[width=1.4cm]{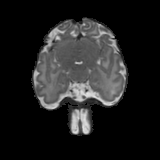}} \hfill
  \subfloat{\includegraphics[width=1.4cm]{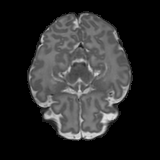}} \hfill
  \subfloat{\includegraphics[width=1.4cm]{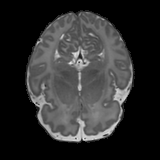}} \hfill
  \subfloat{\includegraphics[width=1.4cm]{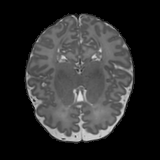}} \hfill
  \subfloat{\includegraphics[width=1.4cm]{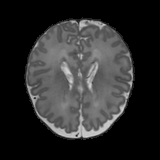}} \hfill
  \subfloat{\includegraphics[width=1.4cm]{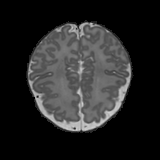}} \hfill
  \subfloat{\includegraphics[width=1.4cm]{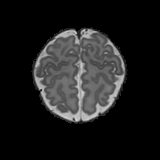}} \\

  \subfloat{\includegraphics[width=1.4cm]{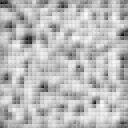}}  \hfill
  \subfloat{\includegraphics[width=1.4cm]{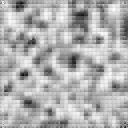}} \hfill
  \subfloat{\includegraphics[width=1.4cm]{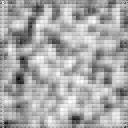}} \hfill
  \subfloat{\includegraphics[width=1.4cm]{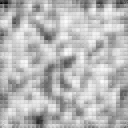}} \hfill
  \subfloat{\includegraphics[width=1.4cm]{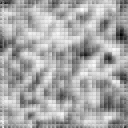}} \hfill
  \subfloat{\includegraphics[width=1.4cm]{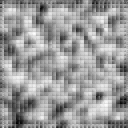}} \hfill
  \subfloat{\includegraphics[width=1.4cm]{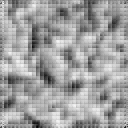}} \hfill
  \subfloat{\includegraphics[width=1.4cm]{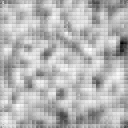}} \\
  
  \setcounter{subfigure}{0}
  \subfloat[$v$: 10]{\includegraphics[width=1.4cm]{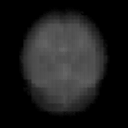}}  \hfill
  \subfloat[$v$: 20]{\includegraphics[width=1.4cm]{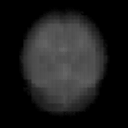}} \hfill
  \subfloat[$v$: 30]{\includegraphics[width=1.4cm]{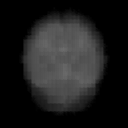}} \hfill
  \subfloat[$v$: 40]{\includegraphics[width=1.4cm]{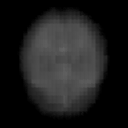}} \hfill
  \subfloat[$v$: 50]{\includegraphics[width=1.4cm]{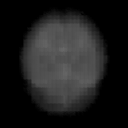}} \hfill
  \subfloat[$v$: 60]{\includegraphics[width=1.4cm]{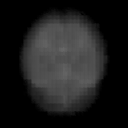}} \hfill
  \subfloat[$v$: 70]{\includegraphics[width=1.4cm]{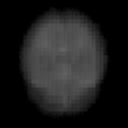}} \hfill
  \subfloat[$v$: 80]{\includegraphics[width=1.4cm]{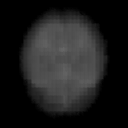}} \\
  
  \caption{Dense sampling the 3D volume using GQN and C-VAE Models. Top Row: Context slices from one particular subject at pose position $v$. Second Row: Ground Truth Images. Third Row: GQN predicted slice at pose position $v$. Bottom Row: C-VAE predicted slice at pose position $v$.}
  \label{fig:gqn-cvae-results}
  \vspace{-7mm} % uncomment for arxiv
\end{figure}

Both models were not able to achieve satisifiable results, as seen in Figure~\ref{fig:gqn-cvae-results}. Due to the complex nature of the brain structure, the GQN model was not able to generalise, and predicted static for all slices. The generated images by the C-VAE model seem to resemble an average of all input context slices with variational noise on top. The experiments also shown that the models do not easily converge, as the latent distribution is far from the prior distribution as measured by the KL divergence. This is notably evident with the GQN, where the KL divergence in the generator module is often very high. Images generated by either method are corrupt.

\subsection{Exp1: ADNI MRI and Thorax CT}

\begin{figure}[ht]
\centering
\iffalse
\resizebox{0.8\textwidth}{!}{%
  \subfloat[Selected slices to use as contexts]{
  \begin{tabular}{@{}ccc@{}}
  \toprule
  \multirow{2}{*}{\begin{tabular}[c]{@{}c@{}}~~\textbf{Number of}~\\ ~~\textbf{Contexts}\end{tabular}}
                & \multicolumn{2}{c}{\textbf{Slice Number, $k$}}                             \\ \cmidrule(l){2-3} 
                & \textbf{ADNI ($K=120$)}           & \textbf{Thorax ($K=100$)}         \\ \midrule
  \textbf{1}    & [60]                              & [50]                              \\
  \textbf{3}    & [40,60,80]                        & [25,50,75]                        \\
  \textbf{5}    & [20,40,60,80,100]                 & [10,30,50,70,90]                  \\
  \textbf{9}    & ~[20,30,40,50,60,70,80,90,100]~   & ~[10,20,30,40,50,60,70,80,90]~~   \\ \bottomrule
  \end{tabular}
  }
} \\
\fi
\resizebox{0.42\textwidth}{!}{%
  \subfloat[ADNI MRI Data-set]{
  \begin{tabular}{@{}ccccccccc@{}}
  \toprule
                  & \multicolumn{4}{c}{~~\textbf{Context Images}~~}   \\ \cmidrule(l){2-5} 
                  & 1         & 3         & 5         & 9             \\ \midrule
  ~~\textbf{SSIM} & ~~0.695~  & ~0.719~   & ~0.733~   & ~0.736~~      \\
  ~~\textbf{CC}   & ~~0.913~  & ~0.917~   & ~0.919~   & ~0.922~~      \\ \bottomrule
  \end{tabular}
  }
} 
\resizebox{0.42\textwidth}{!}{%
  \subfloat[Thorax CT Data-set]{
  \begin{tabular}{@{}ccccccccc@{}}
  \toprule
                  & \multicolumn{4}{c}{~~\textbf{Context Images}~~}   \\ \cmidrule(l){2-5} 
                  & 1         & 3         & 5         & 9             \\ \midrule
  ~~\textbf{SSIM} & ~~0.779~  & ~0.786~   & ~0.787~   & ~0.791~~      \\
  ~~\textbf{CC}   & ~~0.802~  & ~0.815~   & ~0.832~   & ~0.833~~      \\ \bottomrule
  \end{tabular}
  }
} \\ 
% \resizebox{0.9\textwidth}{!}{%
  \subfloat{\includegraphics[width=0.24\textwidth]{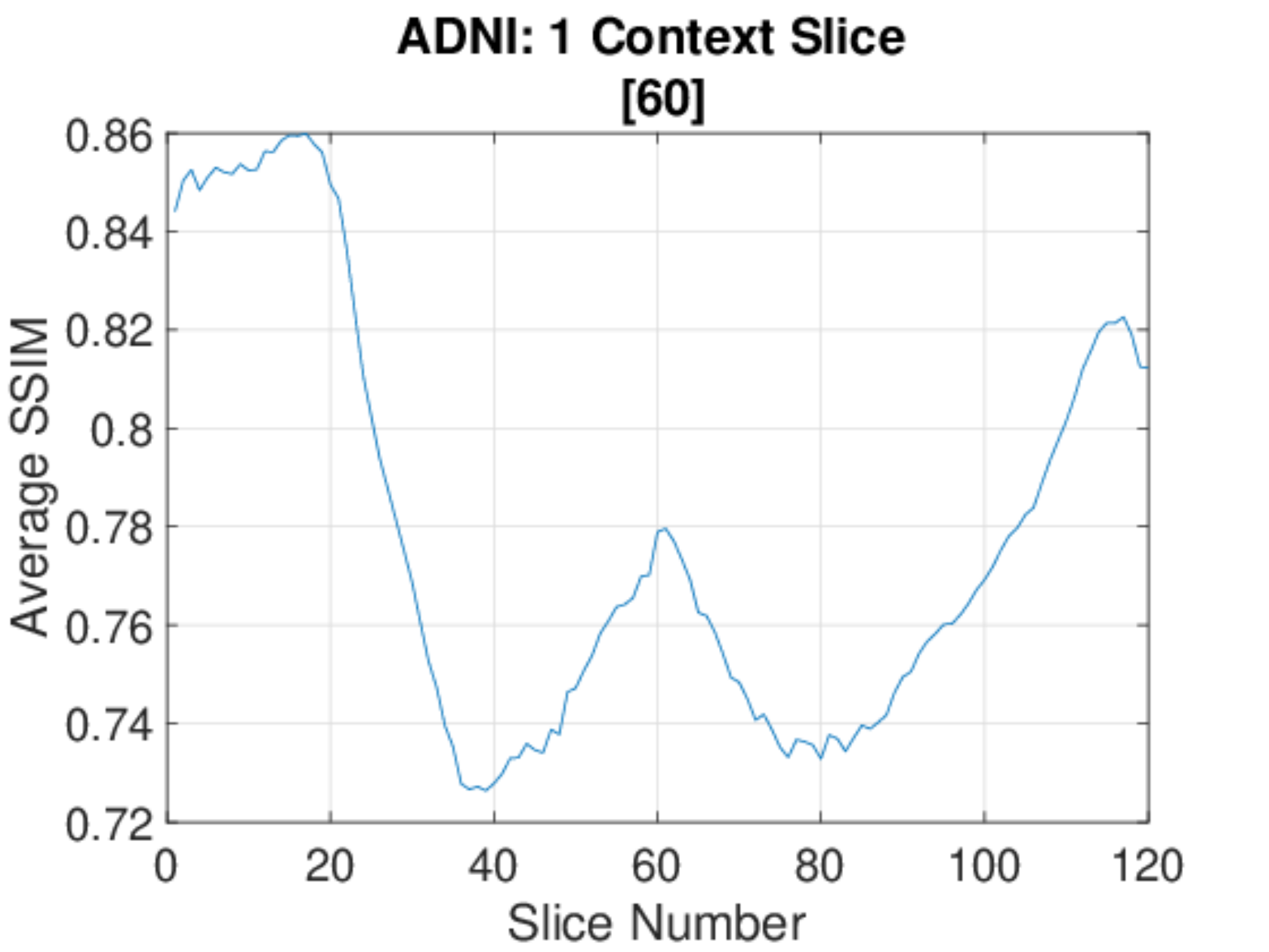}} 
  \subfloat{\includegraphics[width=0.24\textwidth]{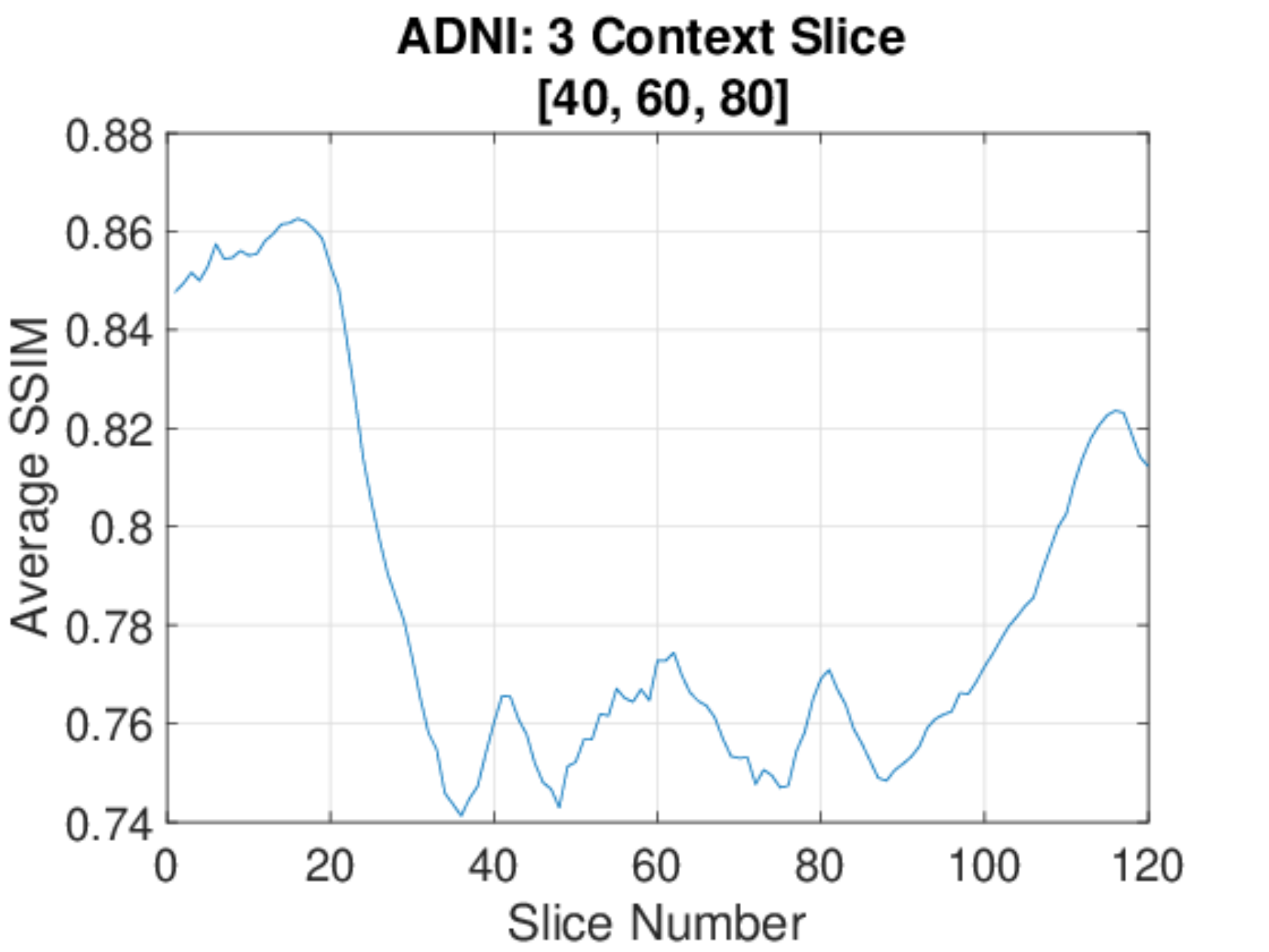}} 
  \subfloat{\includegraphics[width=0.24\textwidth]{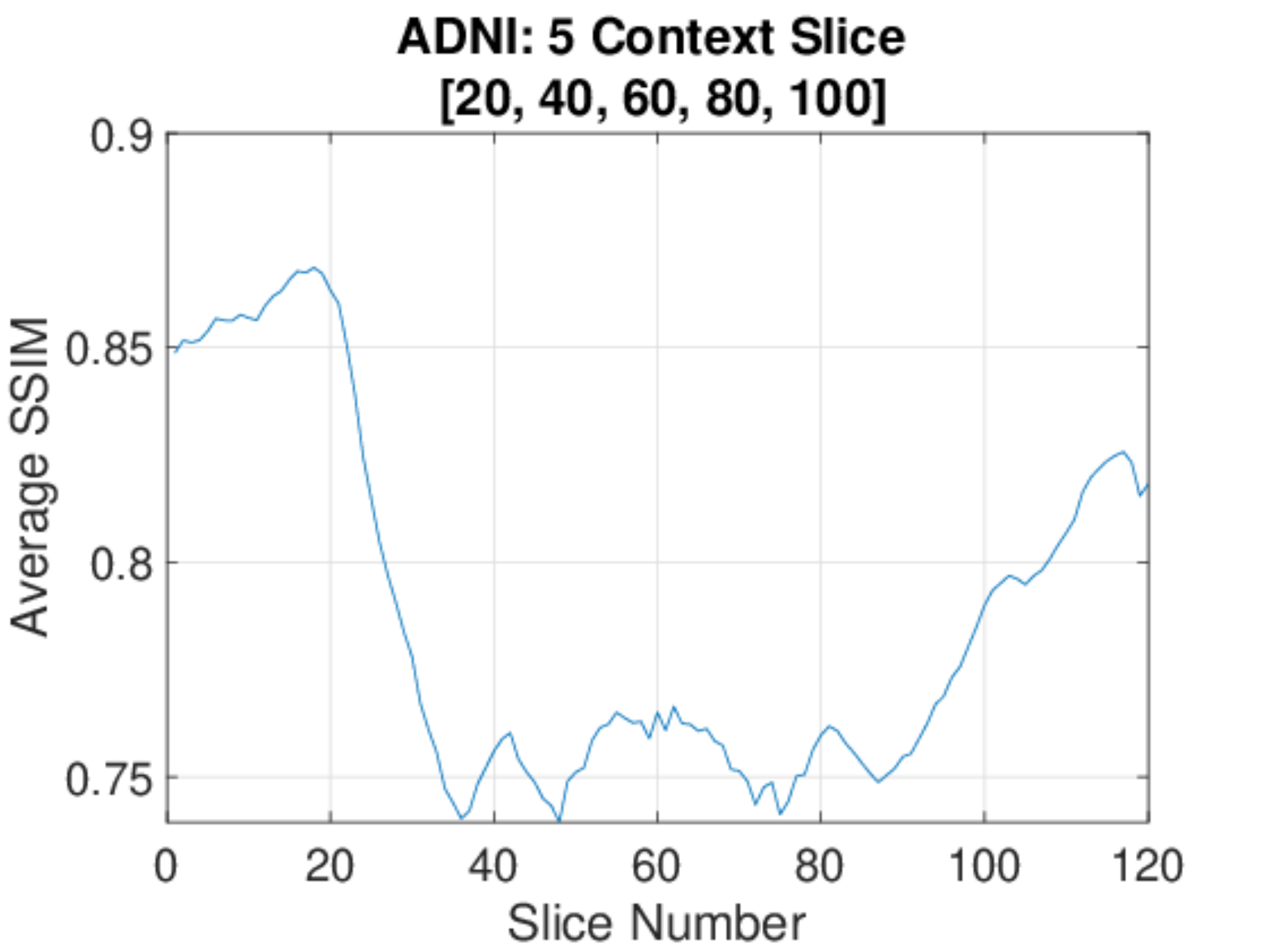}} 
  \subfloat{\includegraphics[width=0.24\textwidth]{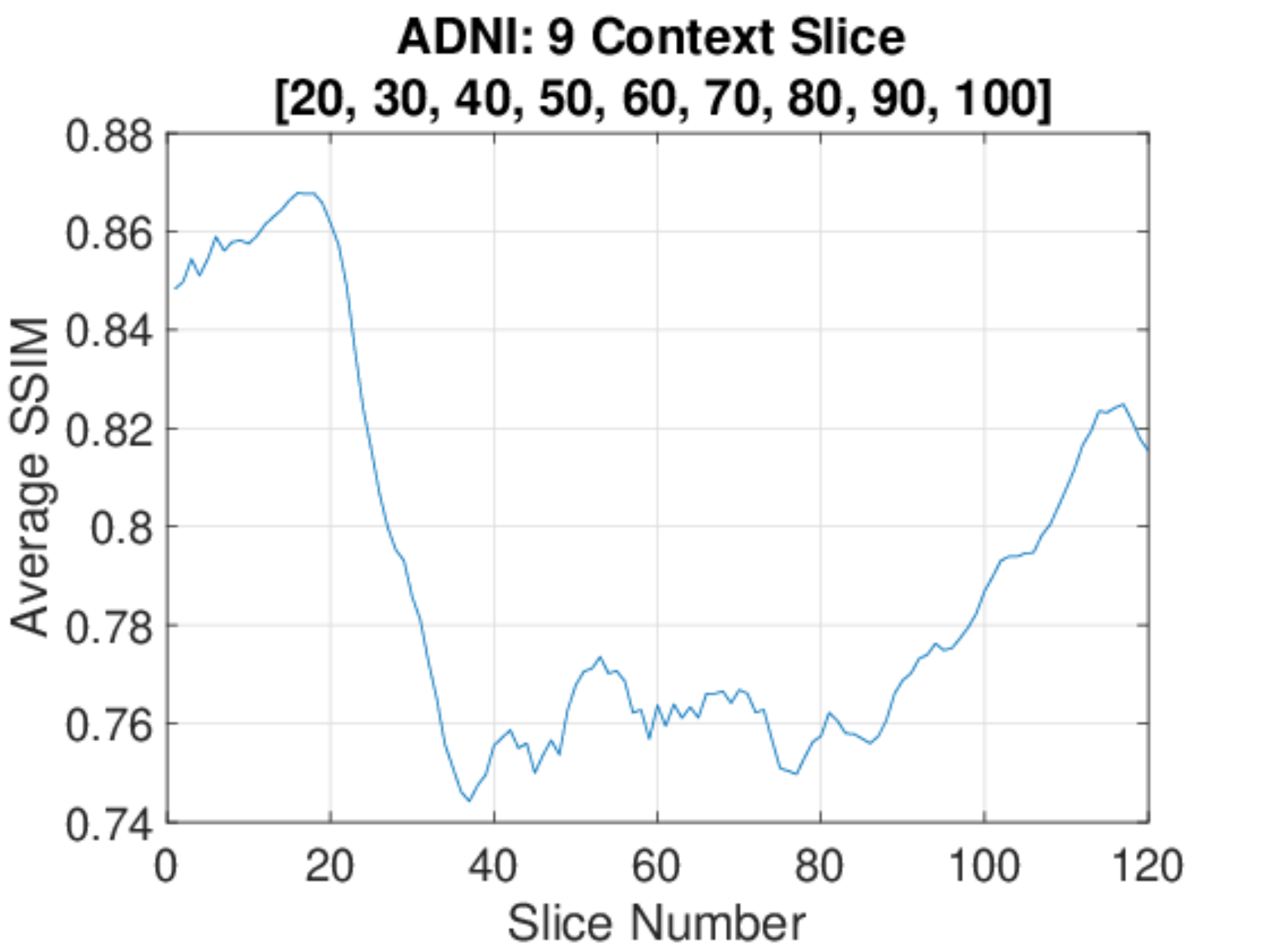}} \\
  \subfloat{\includegraphics[width=0.24\textwidth]{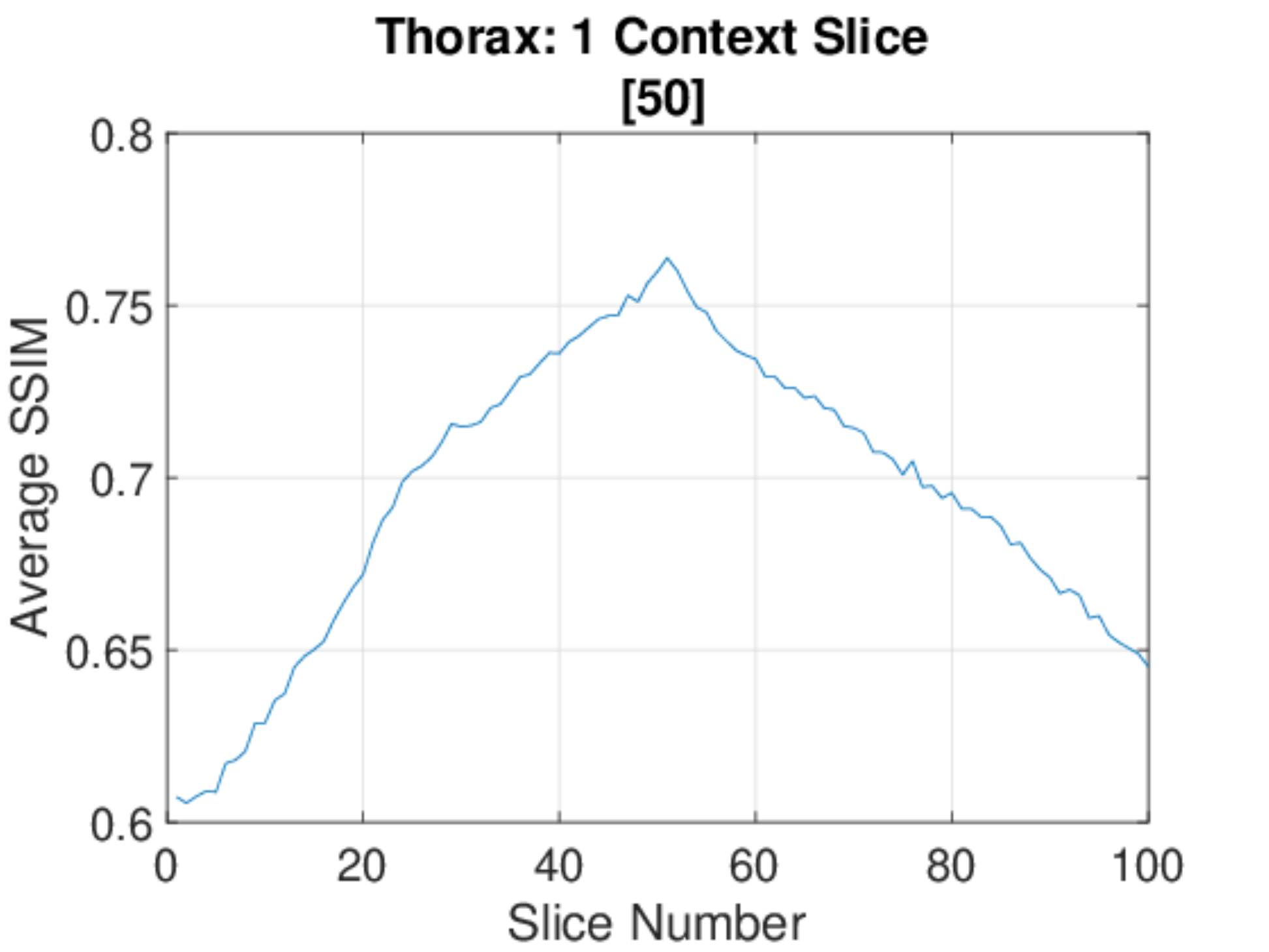}} 
  \subfloat{\includegraphics[width=0.24\textwidth]{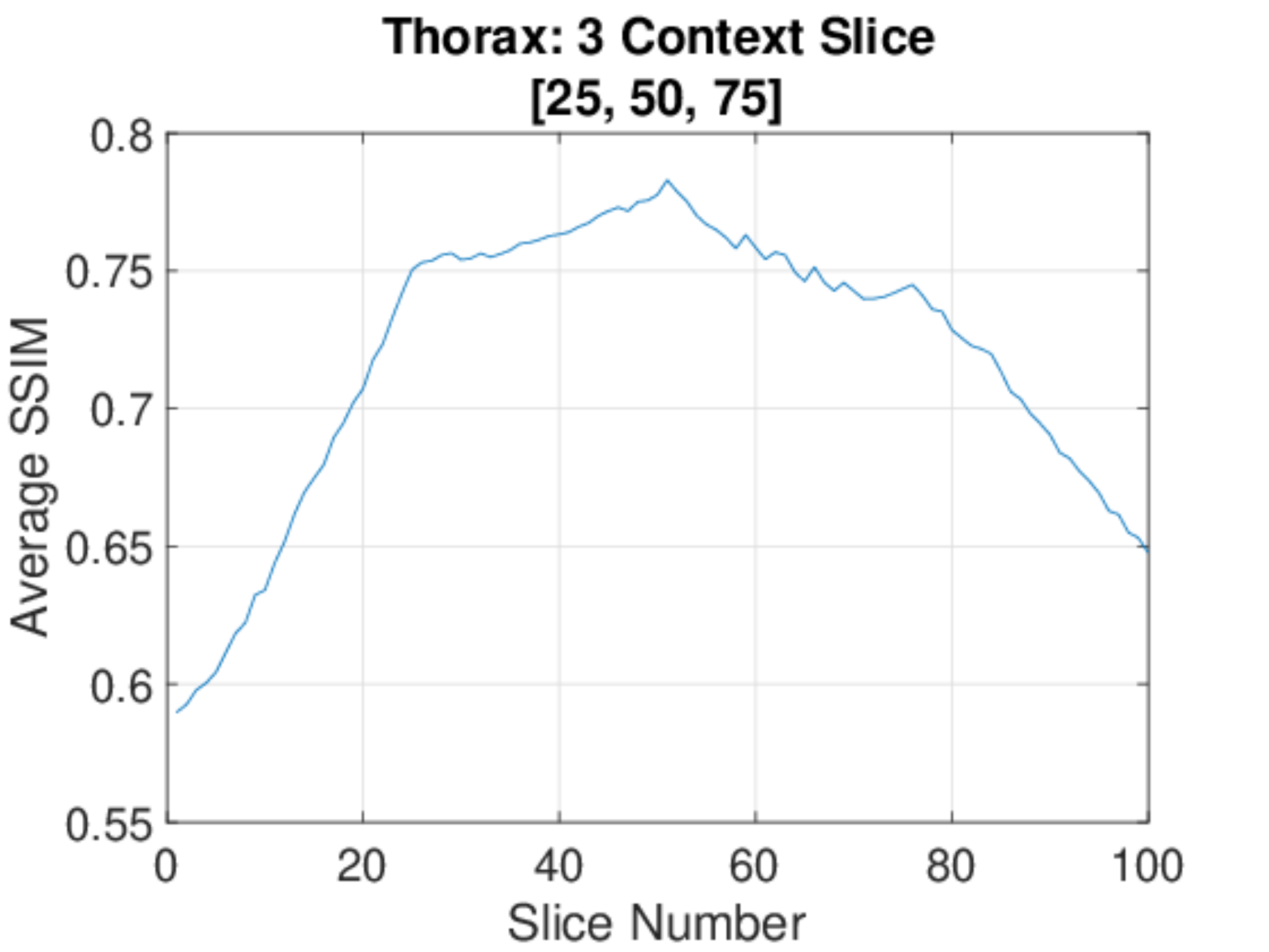}} 
  \subfloat{\includegraphics[width=0.24\textwidth]{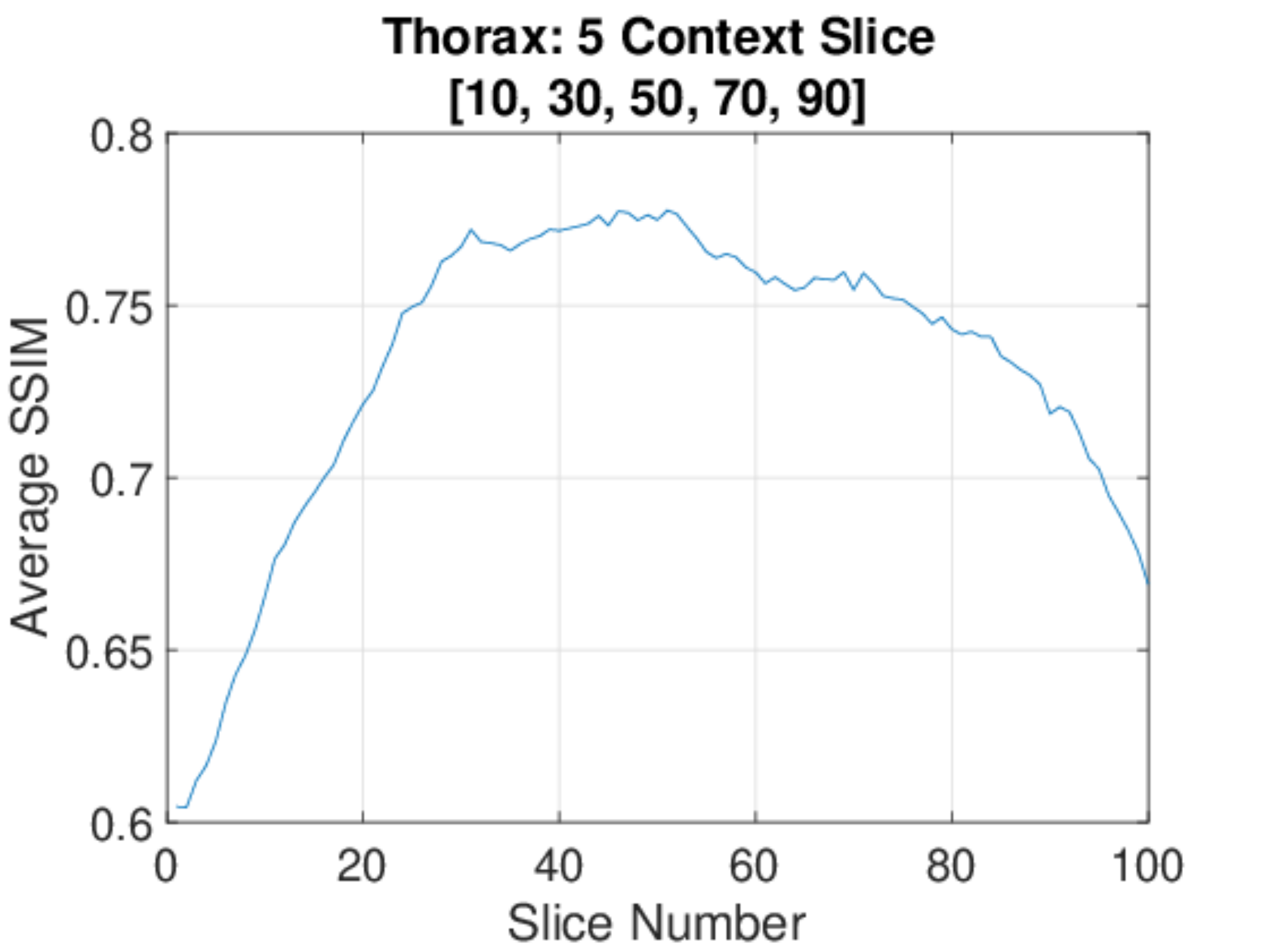}} 
  \subfloat{\includegraphics[width=0.24\textwidth]{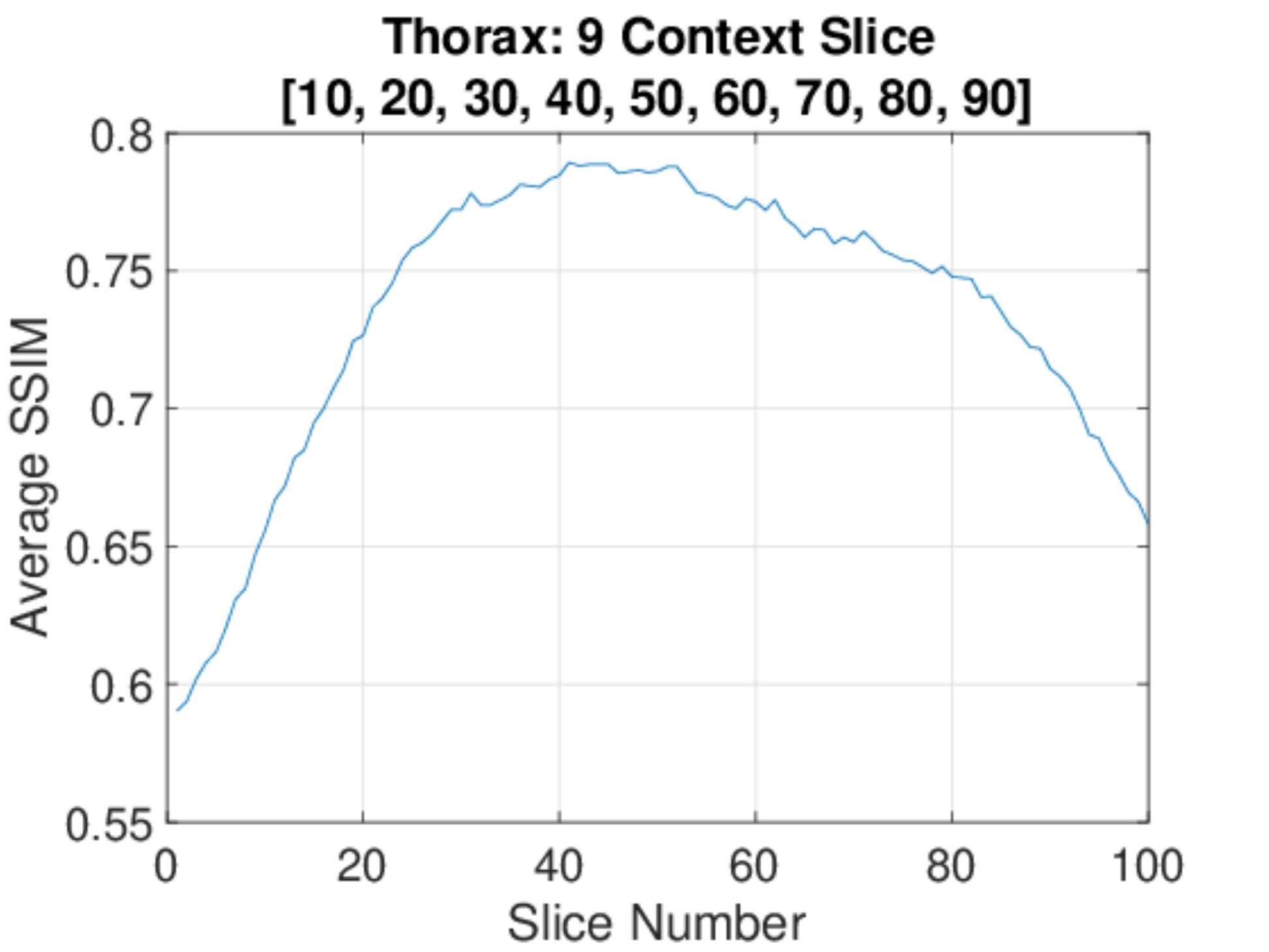}} \\
  (c) Average SSIM vs Slice Number w.r.t. Number of Context Images \\
    Top Row: ADNI MRI Data-set, Bottom Row: Thorax CT Data-set
% } 
\caption{Table and Figures for the Results of Experiment 1}
\label{tab:adni-thorax-benchmark}
\end{figure}

The first set of experiments were used to evaluate the performance of the BRUNO architecture. Tables~\ref{tab:adni-thorax-benchmark}a and \ref{tab:adni-thorax-benchmark}b shows the average SSIM and Cross-Correlation of the generated dense sampled volume compared to the original high resolution volume across all test subjects. 4 experiments were ran with increasing number of contexts; 1, 3, 5 and 9. The contexts selected are sparsely spread to maximise the coverage across the dense volume. In Table~\ref{tab:adni-thorax-benchmark}a and \ref{tab:adni-thorax-benchmark}b it can be seen that as the number of contexts increases, the reconstructed volume becomes closer to the ground truth volume in similarity. 

% \usepackage{booktabs}
% \usepackage{multirow}

Figure~\ref{tab:adni-thorax-benchmark}c shows the average SSIM of each slice across all test subjects for ADNI and Thorax data-sets. A distinct peak in SSIM is perceivable where a query slice aligns with a supplied context slice. This confirms the notion that conditional contexts correctly steer the posterior to a particular part of the distribution. High SSIM for ADNI 0 to 20 and 100 to 120 are the edge cases, where a majority of the content are background. The inverse is the case for thorax, where slices approaching top and bottom have high variability in structure, thus reducing the SSIM.

\subsection{Exp2: Fetal Brain Template Volume}

The second set of experiments has evaluated the usefulness of the proposed approach for fetal MRI reconstruction: State-of-the-art iterative image-based reconstruction methods, e.g. Slice-to-Volume Reconstruction (SVR)~\cite{kainz2015fast}, often require a good initialisation volume for the initial target registration. In clinical setting, especially in fetal MRI, volumes are often motion corrupted if the fetus is awake and constantly moving during image acquisition. Neighbouring slices of the volumes are therefore incoherent and in disarray. In this experiment, BRUNO is used to create the initial registration target volume for 2D to 3D fetal brain reconstruction. During fetal MRI, a few images are often acquired in parallel (usually four, spatially far apart images, at once). These image batches  can be used as conditional contexts for BRUNO. Due to fast parallel acquisition, the slices can be assumed to be aligned and motion free. 

\begin{figure}[!h]
\centering
\resizebox{0.42\textwidth}{!}{%
\subfloat[Selected Slices as Context]{
\begin{tabular}[b]{@{}cc@{}}
\toprule
\begin{tabular}[c]{@{}c@{}}~~\textbf{Number of}~\\ ~~\textbf{Contexts}\end{tabular} 
    & \begin{tabular}[c]{@{}c@{}}\textbf{Slice Number, $k$}\\ \textbf{($K=80$)}\end{tabular}  \\ \midrule
\textbf{1}   &  {[}40{]}                        \\
\textbf{3}   &  {[}20,40,60{]}                  \\
\textbf{4}   &  {[}10,30,50,70{]}               \\
\textbf{7}   & ~{[}10,20,30,40,50,60.70{]}~~    \\ \bottomrule
\end{tabular}
}
} \quad
\resizebox{0.42\textwidth}{!}{%
\subfloat[SSIM/CC of Reconstructed Volume]{
\begin{tabular}[b]{@{}ccccccccc@{}}
\toprule
                & \multicolumn{4}{c}{~~\textbf{Context Images}~~}       \\ \cmidrule(l){2-5} 
                & 1            & 3           & 4           & 7          \\ \midrule
~~\textbf{SSIM} & ~~0.665~     & ~0.676~     & ~0.679~     & ~0.684~~   \\
~~\textbf{CC}   & ~~0.868~     & ~0.875~     & ~0.876~     & ~0.880~~   \\ \bottomrule
\end{tabular}
}
} \\
\subfloat[Average SSIM vs Slice Number w.r.t. Number of Context Images]{
  \includegraphics[width=0.23\textwidth]{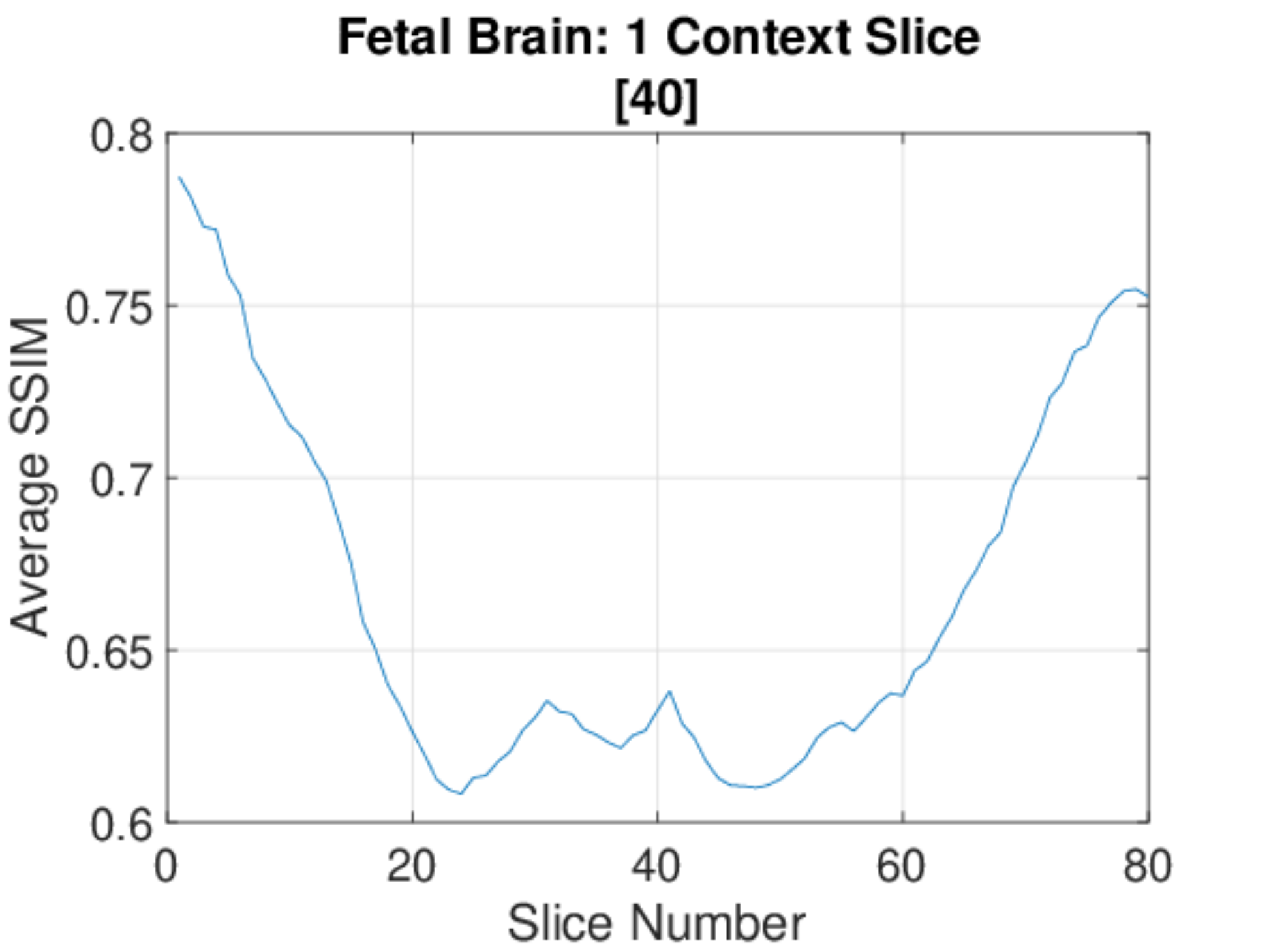} 
  \includegraphics[width=0.23\textwidth]{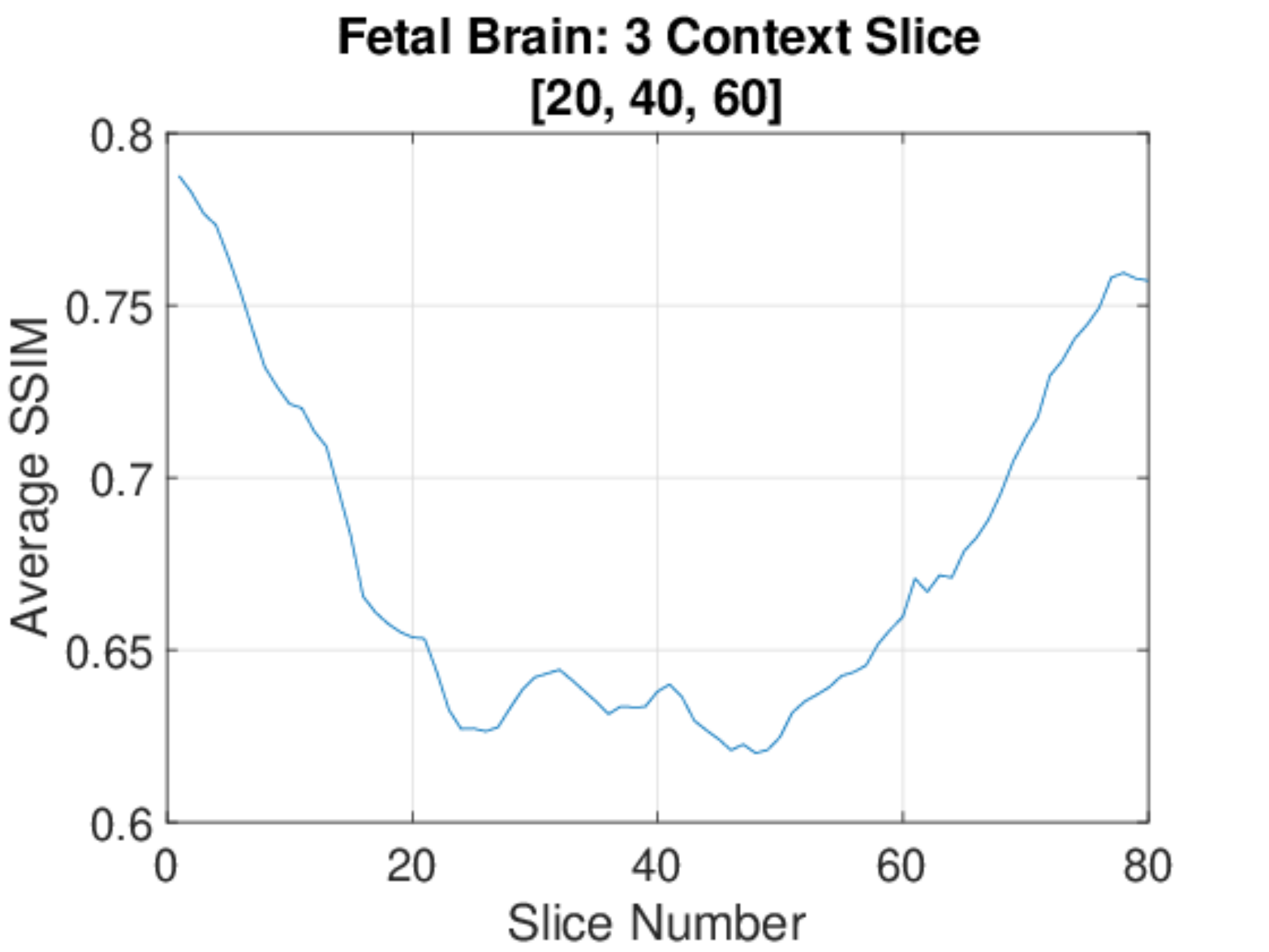} 
  \includegraphics[width=0.23\textwidth]{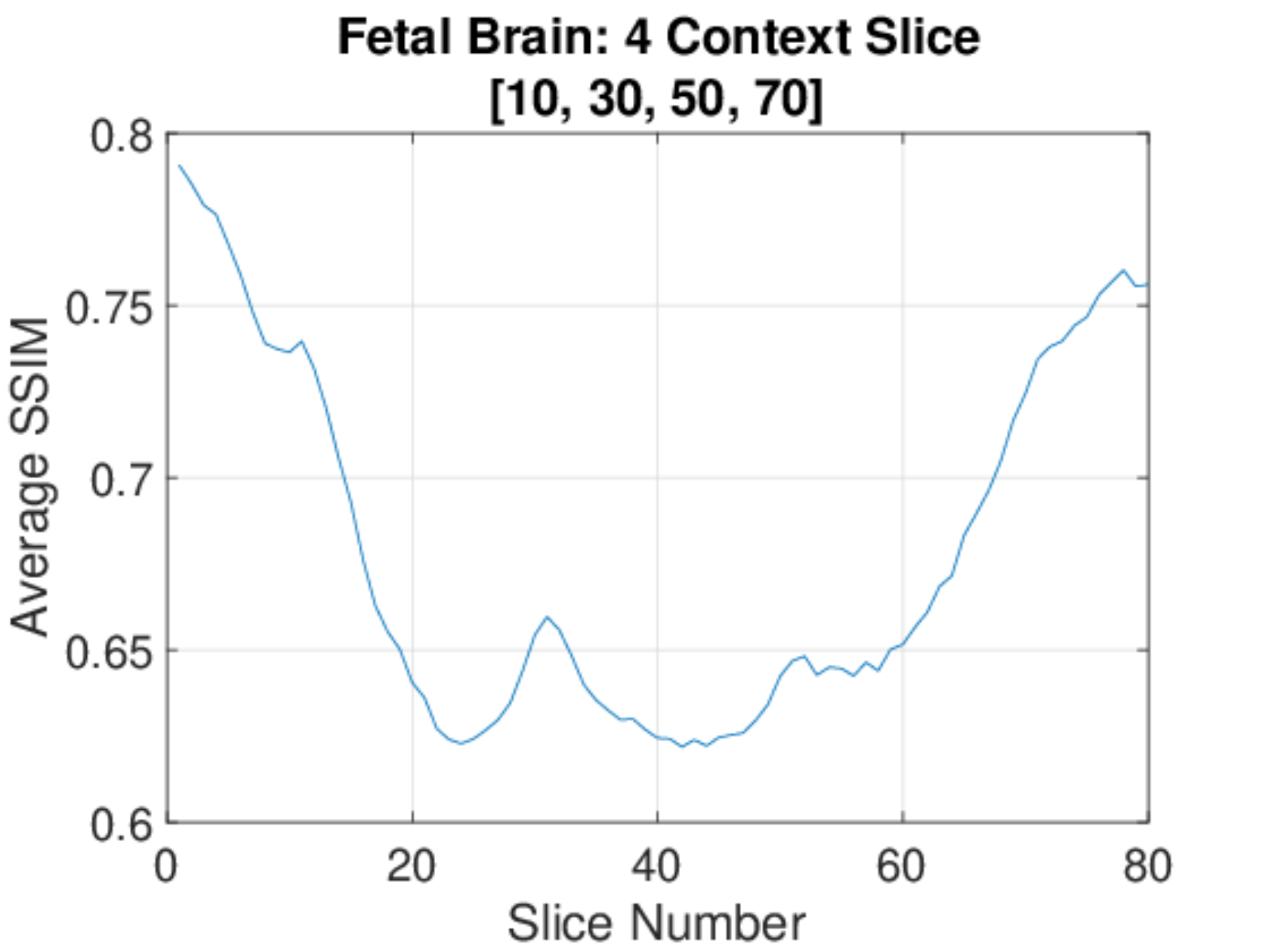} 
  \includegraphics[width=0.23\textwidth]{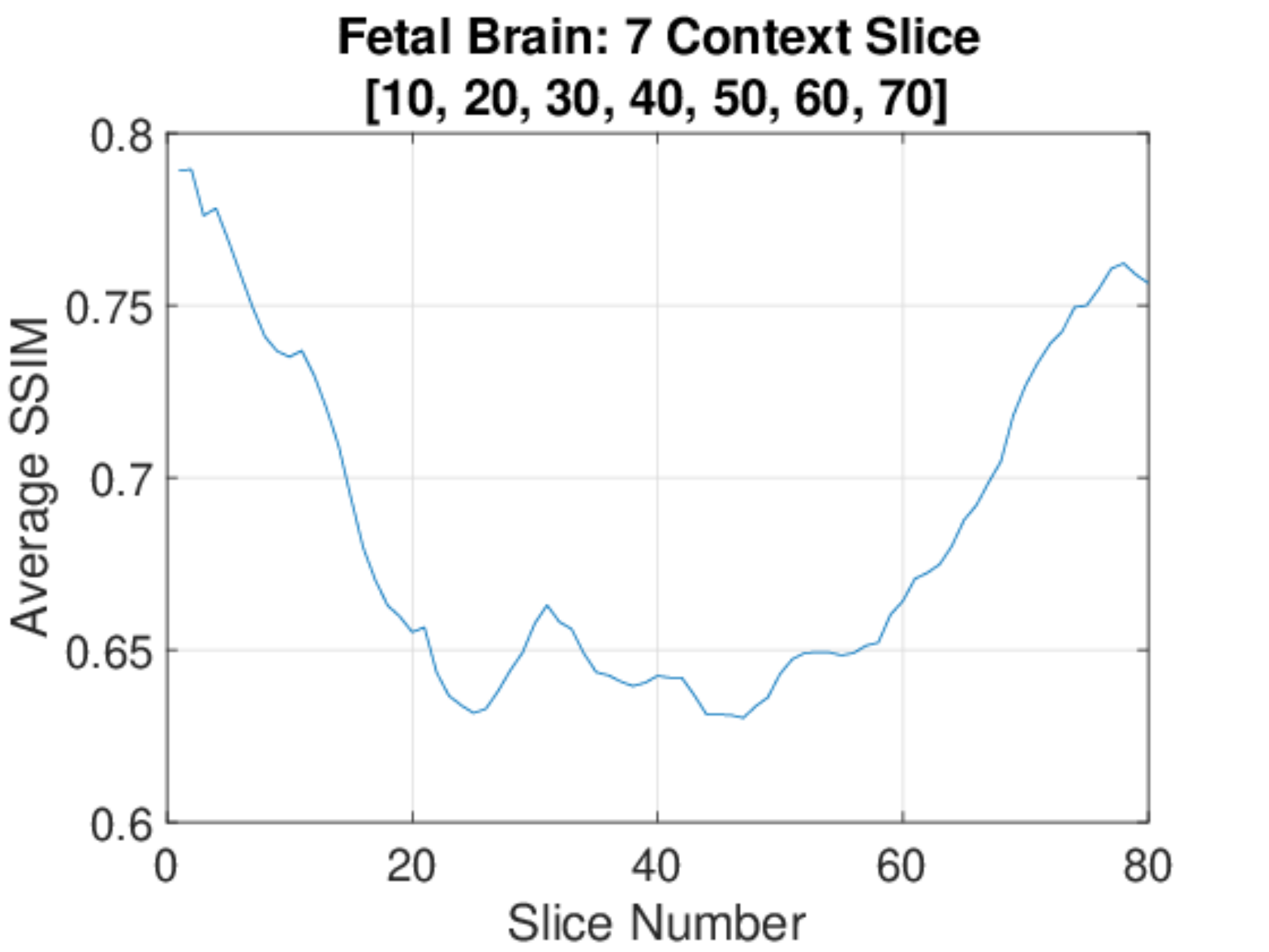}
}
\caption{Table and Figures for the Results of Experiment 2}
\label{tab:fetal-benchmark}
\end{figure}

As with the first set of experiments, the performance of BRUNO is evaluated with varying context slices. In total, there are 80 slices in the dense fetal brain volume. Table~\ref{tab:fetal-benchmark}a below shows the number of contexts, with the corresponding slice numbers, that is used during inference. Table~\ref{tab:fetal-benchmark}b shows the SSIM and CC of average reconstructed SVR initialisation volumes. Like as in the first experiments, as the number of context images increase, the average SSIM and Cross-Correlation increases. Distinct peaks in SSIM are also present where a query slice approaches a supplied context slice.

Figure~\ref{fig:neonatal-examples} below shows more examples of samples drawn from the posterior distribution that have been conditioned by the 4 context slices. 

\begin{figure}[!h]
  \centering
  \subfloat[$v$: 10]{\includegraphics[width=1.4cm]{CC00388XX18_1.png}} \hspace{2mm}
  \subfloat[$v$: 30]{\includegraphics[width=1.4cm]{CC00388XX18_2.png}} \hspace{2mm}
  \subfloat[$v$: 50]{\includegraphics[width=1.4cm]{CC00388XX18_3.png}} \hspace{2mm}
  \subfloat[$v$: 70]{\includegraphics[width=1.4cm]{CC00388XX18_4.png}} \hspace{2mm} \\
  
  \subfloat{\includegraphics[width=1.4cm]{gt_img0.png}}  \hfill
  \subfloat{\includegraphics[width=1.4cm]{gt_img10.png}} \hfill
  \subfloat{\includegraphics[width=1.4cm]{gt_img20.png}} \hfill
  \subfloat{\includegraphics[width=1.4cm]{gt_img30.png}} \hfill
  \subfloat{\includegraphics[width=1.4cm]{gt_img40.png}} \hfill
  \subfloat{\includegraphics[width=1.4cm]{gt_img50.png}} \hfill
  \subfloat{\includegraphics[width=1.4cm]{gt_img60.png}} \hfill
  \subfloat{\includegraphics[width=1.4cm]{gt_img70.png}} \\
  
  \subfloat{\includegraphics[width=1.4cm]{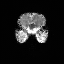}}  \hfill
  \subfloat{\includegraphics[width=1.4cm]{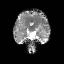}} \hfill
  \subfloat{\includegraphics[width=1.4cm]{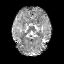}} \hfill
  \subfloat{\includegraphics[width=1.4cm]{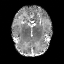}} \hfill
  \subfloat{\includegraphics[width=1.4cm]{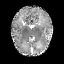}} \hfill
  \subfloat{\includegraphics[width=1.4cm]{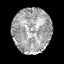}} \hfill
  \subfloat{\includegraphics[width=1.4cm]{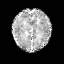}} \hfill
  \subfloat{\includegraphics[width=1.4cm]{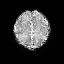}} \\ 

  \subfloat{\includegraphics[width=1.4cm]{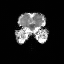}}  \hfill 
  \subfloat{\includegraphics[width=1.4cm]{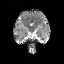}} \hfill 
  \subfloat{\includegraphics[width=1.4cm]{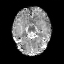}} \hfill 
  \subfloat{\includegraphics[width=1.4cm]{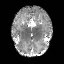}} \hfill
  \subfloat{\includegraphics[width=1.4cm]{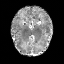}} \hfill 
  \subfloat{\includegraphics[width=1.4cm]{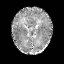}} \hfill 
  \subfloat{\includegraphics[width=1.4cm]{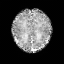}} \hfill 
  \subfloat{\includegraphics[width=1.4cm]{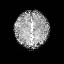}} \\ 

  \subfloat{\includegraphics[width=1.4cm]{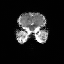}}  \hfill 
  \subfloat{\includegraphics[width=1.4cm]{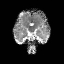}} \hfill 
  \subfloat{\includegraphics[width=1.4cm]{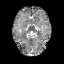}} \hfill 
  \subfloat{\includegraphics[width=1.4cm]{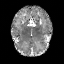}} \hfill
  \subfloat{\includegraphics[width=1.4cm]{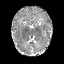}} \hfill 
  \subfloat{\includegraphics[width=1.4cm]{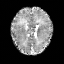}} \hfill 
  \subfloat{\includegraphics[width=1.4cm]{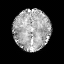}} \hfill 
  \subfloat{\includegraphics[width=1.4cm]{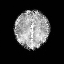}} \\ 

  \subfloat{\includegraphics[width=1.4cm]{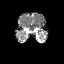}}  \hfill 
  \subfloat{\includegraphics[width=1.4cm]{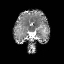}} \hfill 
  \subfloat{\includegraphics[width=1.4cm]{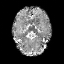}} \hfill 
  \subfloat{\includegraphics[width=1.4cm]{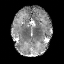}} \hfill
  \subfloat{\includegraphics[width=1.4cm]{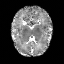}} \hfill 
  \subfloat{\includegraphics[width=1.4cm]{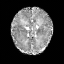}} \hfill 
  \subfloat{\includegraphics[width=1.4cm]{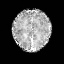}} \hfill 
  \subfloat{\includegraphics[width=1.4cm]{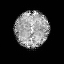}} \\ 

  \subfloat{\includegraphics[width=1.4cm]{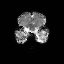}}  \hfill 
  \subfloat{\includegraphics[width=1.4cm]{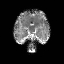}} \hfill 
  \subfloat{\includegraphics[width=1.4cm]{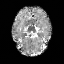}} \hfill 
  \subfloat{\includegraphics[width=1.4cm]{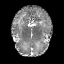}} \hfill
  \subfloat{\includegraphics[width=1.4cm]{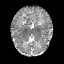}} \hfill 
  \subfloat{\includegraphics[width=1.4cm]{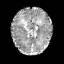}} \hfill 
  \subfloat{\includegraphics[width=1.4cm]{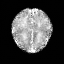}} \hfill 
  \subfloat{\includegraphics[width=1.4cm]{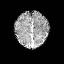}} \\ 

  \setcounter{subfigure}{0}
  \subfloat[$v$:  0]{\includegraphics[width=1.4cm]{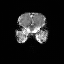}}   \hfill
  \subfloat[$v$: 10]{\includegraphics[width=1.4cm]{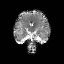}} \hfill 
  \subfloat[$v$: 20]{\includegraphics[width=1.4cm]{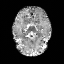}} \hfill 
  \subfloat[$v$: 30]{\includegraphics[width=1.4cm]{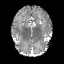}} \hfill
  \subfloat[$v$: 40]{\includegraphics[width=1.4cm]{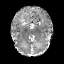}} \hfill 
  \subfloat[$v$: 50]{\includegraphics[width=1.4cm]{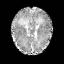}} \hfill 
  \subfloat[$v$: 60]{\includegraphics[width=1.4cm]{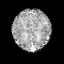}} \hfill 
  \subfloat[$v$: 70]{\includegraphics[width=1.4cm]{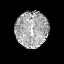}} \\ 

  \caption{A Conditional BRUNO architecture, consisting of a trainable bijector and Student's t-process statistical modelling. Top Row: Context slices from one particular test subject at pose position $v$, Second Row: Ground Truth Images, Third Row On-wards: six predicted slices at pose position $v$ sampled from the posterior distribution conditioned by the contexts.}
  \label{fig:neonatal-examples}
\end{figure}

\begin{figure}[ht]
  \centering
  \includegraphics[width=0.9\textwidth]{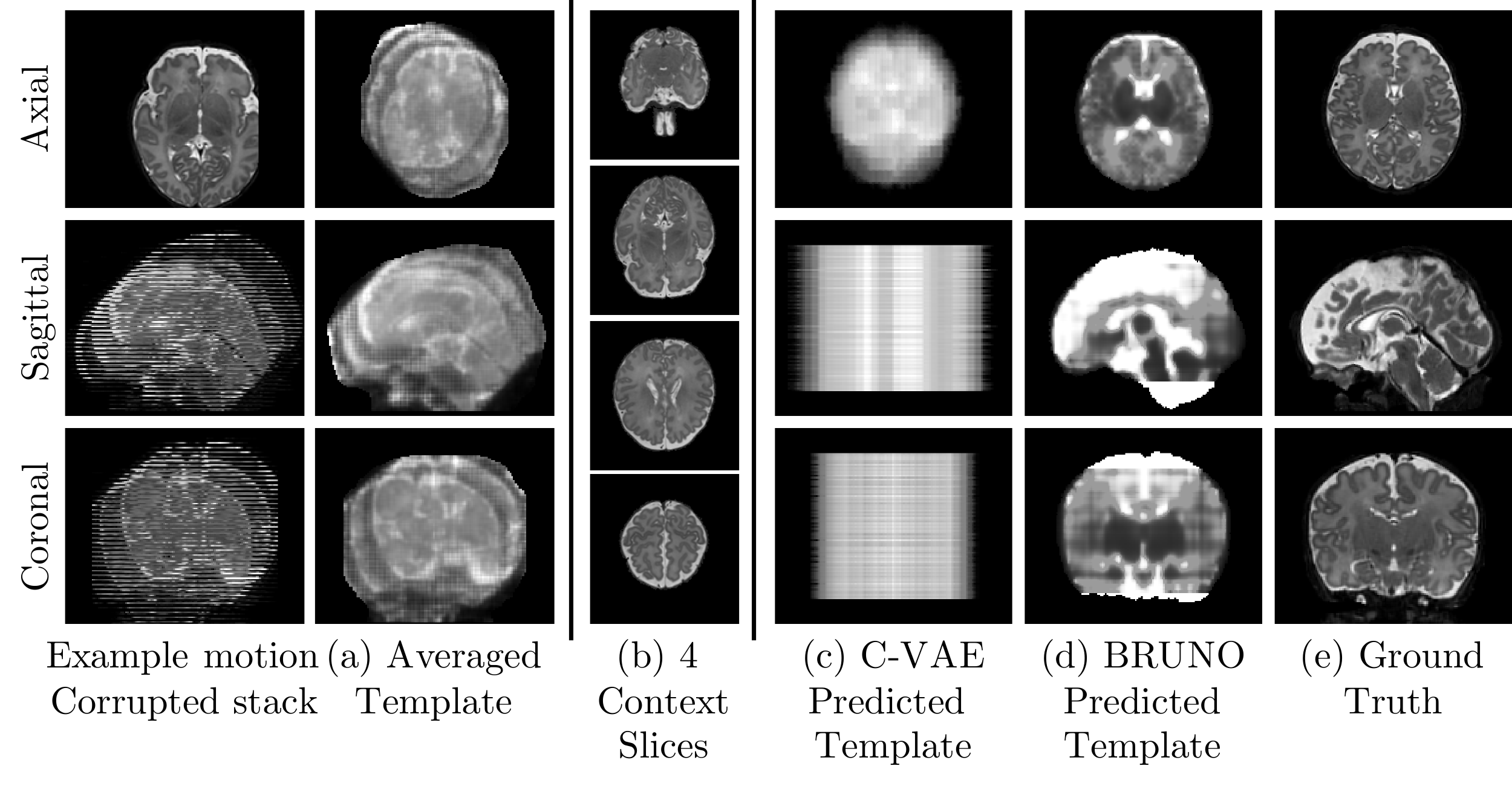}
  \caption{For motion compensation of fetal and neonatal MRI, averaged template volumes (a) from all acquired slices (several 100) are used to initialise SVR~\cite{kainz2015fast}. Only four images were used (b) from an image batch that has been acquired in parallel, (common in fetal and neonatal MRI, i.e. without motion corruption between the four images). A C-VAE is not able to predict a template volume from these four images (c), while our approach using BRUNO predicts a reasonable volume (d) compared to the ground truth (e).    }
  \label{fig:bruno-reconstruction-compare}
\end{figure}

Fetuses can move up to 20mm~\cite{kainz2015fast} when active. For SVR, a Gaussian average is taken of all acquired stacks to be used as the initial registration target. If motion is too severe this initialisation volume will not be sufficient for subsequent iterative reconstruction. To simulate motion, high resolution reconstructed volumes are synthetically motion corrupted to be used as a base line. Three orthogonal stacks are made with 10mm of random motion in translation only. On average, the SSIM of Gaussian averaged volumes motion corrupted volumes compared to the original high resolution volume is 0.297. Depending on the robustness of the SVR algorithm used, reconstruction may, or may not, be possible. A BRUNO generated volume is made by densely and repeatedly sampling all possible pose positions, with the final volume being an average of all sampled slices. Only four context images are used. The generated volume was able to achieve an SSIM of 0.679 compared to the original high resolution volume, this is shown in Figure~\ref{fig:bruno-reconstruction-compare}.

\section{Conclusion and Discussion}

This paper introduces the idea of using Deep Neural Networks and Bayesian Deep Learning to build mental maps of anatomy of various medical volumes. A conditional generative model, based on the BRUNO architecture, is trained on existing high resolution 3D volumes. It can  be used to create patient specific volumes by densely querying all possible pose positions, whilst conditioned by a few existing slices that are used as contexts. \emph{Exp1} shows that BRUNO is able to reconstruct MRI brain volumes with an SSIM of 0.7 and Thorax volumes with an SSIM of 0.8  compared to the original high resolution ground truth. \emph{Exp2} shows a specific use case for BRUNO to generate initial target volumes for 2D to 3D fetal brain MRI reconstruction. 

Future work will be to investigate further into the framework, and improve image quality generation as well as to introduce more Degrees-of-Freedom (DoF). In the current implementation, BRUNO is able to successfully traverse single DoF, and is applicable for use cases such CT and MRI. Increased DoF, with added rotations and translations, can be particularly valuable for modalities such as freehand ultrasound, with applications for Reinforcement Learning and 3D scene exploration. 

\subsection{Acknowledgements}
We thank The Wellcome Trust IEH Award iFind project [102431], Innovate UK: London Medical Imaging \& Artificial Intelligence Centre for Value-Based Healthcare [104691], and NVIDIA for their GPU donations. The data used in the preparation of this article were obtained from the Alzheimer's Disease Neuroimaging Initiative (ADNI) database (\url{http://adni.loni.usc.edu}). Fetal brain data were accessed only with informed consent, subject to approval and  formal Data Sharing Agreement. We also like to thank Ira, author of BRUNO and Conditional BRUNO, for the valuable discussions.

\bibliographystyle{splncs03}
\bibliography{references}

\begin{thebibliography}{10}
\providecommand{\url}[1]{\texttt{#1}}
\providecommand{\urlprefix}{URL }

\bibitem{cerrolaza20183d}
Cerrolaza, J.J., et~al.: 3d fetal skull reconstruction from 2dus via deep
  conditional generative networks. In: MICCAI'18. pp. 383--391. Springer (2018)

\bibitem{choy20163d}
Choy, C.B., et~al.: {3D-R2N2: A Unified Approach for Single and Multi-view 3D
  Object Reconstruction}. In: ECCV'16. pp. 628--644. Springer (2016)

\bibitem{10.1007/11569541_46}
Ding, F., et~al.: {Segmentation of 3D CT volume images using a singFast
  generation of virtual X-ray images for reconstruction of 3D anatomyle 2D
  atlas}. In: International Workshop on Computer Vision for Biomedical Image
  Applications. pp. 459--468. Springer (2005)

\bibitem{DBLP:journals/corr/DinhSB16}
Dinh, L., et~al.: Density estimation using real {NVP}. CoRR  abs/1605.08803
  (2016)

\bibitem{ehlke2013fast}
Ehlke, M., et~al.: Fast generation of virtual x-ray images for reconstruction
  of 3d anatomy. IEEE transactions on visualization and computer graphics
  19(12),  2673--2682 (2013)

\bibitem{Eslami1204}
Eslami, S.M.A., et~al.: Neural scene representation and rendering. Science
  360(6394),  1204--1210 (2018),
  \url{https://science.sciencemag.org/content/360/6394/1204}

\bibitem{goodfellow2014generative}
Goodfellow, I., et~al.: Generative adversarial nets. In: NIPS. pp. 2672--2680
  (2014)

\bibitem{ogroth_2019}
Groth, O.: ogroth/tf-gqn (Jun 2019), \url{https://github.com/ogroth/tf-gqn}

\bibitem{kainz2015fast}
Kainz, B., et~al.: Fast volume reconstruction from motion corrupted stacks of
  2d slices. IEEE transactions on medical imaging  34(9),  1901--1913 (2015)

\bibitem{DBLP:journals/corr/KingmaW13}
Kingma, D.P., et~al.: Auto-encoding variational bayes. CoRR  abs/1312.6114
  (2013)

\bibitem{korshunovaconditional}
Korshunova, I., et~al.: Conditional bruno: A deep recurrent process for
  exchangeable labelled data. Bayesian Deep Learning NeurIPS Workshop  (2018)

\bibitem{Kunter2009}
Kunter, M., et~al.: {Unsupervised object segmentation for 2D to 3D conversion}.
  In: Stereoscopic Displays and Applications XX. vol. 7237, p. 72371B (2009)

\bibitem{papamakarios2017masked}
Papamakarios, G., Pavlakou, T., Murray, I.: Masked autoregressive flow for
  density estimation. In: NIPS. pp. 2338--2347 (2017)

\bibitem{rezende2015variational}
Rezende, D.J., et~al.: Variational inference with normalizing flows. In:
  {ICML}. {JMLR} Workshop and Conference Proceedings, vol.~37, pp. 1530--1538.
  JMLR.org (2015)

\bibitem{ronneberger2015u}
Ronneberger, O., et~al.: U-net: Convolutional networks for biomedical image
  segmentation. In: MICCAI'15. pp. 234--241. Springer (2015)

\bibitem{DBLP:conf/nips/SohnLY15}
Sohn, K., et~al.: {Learning Structured Output Representation using Deep
  Conditional Generative Models}. In: {NeurIPS}. pp. 3483--3491 (2015)

\bibitem{wang2017unsupervised}
Wang, L., et~al.: Unsupervised 3d reconstruction from a single image via
  adversarial learning. CoRR  abs/1711.09312 (2017)

\bibitem{DBLP:conf/nips/ZaheerKRPSS17}
Zaheer, M., et~al.: {Deep Sets}. In: {NeurIPS}. pp. 3394--3404 (2017)

\end{thebibliography}

\end{document}